\documentclass[]{interact}

\usepackage{makecell}
\usepackage{epstopdf}
\usepackage{subfigure}

\theoremstyle{plain}

\theoremstyle{definition}

\theoremstyle{remark}

\usepackage{bbm}
\usepackage{setspace}
\usepackage{pdfpages}
\usepackage{amsmath}
\usepackage{hyperref}
\hypersetup{
	colorlinks   = true, 
	urlcolor     = blue, 
	linkcolor    = blue, 
	citecolor   = blue 
}

\usepackage{graphicx}

\begin{document}
\newcommand{\bX}{\mbox {\bf X}}
\newcommand{\bI}{\mbox {\bf I}}
\newcommand{\bgamma}{\mbox {\bf \gamma}}
\newcommand{\balpha}{\mbox {\bf \alpha}}
\newcommand{\bdelta}{\mbox {\bf \delta}}
\newcommand{\bbeta}{\mbox {\bf \beta}}
\newcommand{\bone}{\mbox {\bf 1}}
\newcommand{\nn}{\nonumber}
\newcommand{\beqn}{\begin{equation}}
\newcommand{\eeqn}{\end{equation}}

\title{Weighting Based Approaches to Borrowing Historical Controls for Indirect comparison for Time-to-Event Data with a Cure Fraction}
\author{Jixian Wang\textsuperscript{a},  Hongtao Zhang\textsuperscript{b}, Ram Tiwari\textsuperscript{c}\\
\textsuperscript{a}GBDS, Bristol Myers Squibb, Boudry, Switzerland;\\
\textsuperscript{b}Biostatistics and Research Decision Sciences, Merck \& Co., Inc., North Wales, Pennsylvania, USA;\\
\textsuperscript{c}GBDS, Bristol Myers Squibb, Berkeley Heights, New Jersey, USA}

\maketitle

\begin{abstract}
To use historical controls for indirect comparison with single-arm trials, the population difference between data sources should be adjusted to reduce confounding bias. The adjustment is more difficult for time-to-event data with a cure fraction. We propose different adjustment approaches based on pseudo observations and calibration weighting by entropy balancing. We show a simple way to obtain the pseudo observations for the cure rate and propose a simple weighted estimator based on them.  Estimation of the survival function in presence of a cure fraction is also considered. Simulations are conducted to examine the proposed approaches. An application to a breast cancer study is presented. 
\end{abstract}


\section{Introduction}

Recently, real-world data (RWD) have been increasingly utilized to establish historical controls for single-arm studies or to  augment the control and/or treated arm in a  small randomized controlled trials. Here, we mainly focus on indirect comparison, that is,  using historical data as the only control subjects, although our approaches can also be used to augment the control group of an RCT with historical data.
The trial and RWD populations are often very different. Consequently, the differences between the two populations may lead to confounding biases in statistical inference, such as the indirect comparisons between the treated arm and the historical control. Several adjustment methods are available that can be used to minimize confounding bias. Typical ones include direct adjustment based on the outcome regression model, inverse probability weighting (IPW) and propensity score matching (PSM) (Rosenbaum \& Rubin, 1983; Lunceford \& Davidian, 2004). The latter two match or balance the chance of being treated; hence, the weighted or matched data may be considered as having treatment quasi-randomized.   More recently developed calibration weighting method aims at balancing the covariate summary statistics, rather than propensity scores, based on calibration estimation (CE) in survey sampling (Deville \& S\"arndal,  1992; Devaud \& Tillé, 2019). The matching-adjusted indirect comparison (MAIC) (Signorovitc et al. 2012), widely used in health technology assessment (HTA), is a special case of CE. The MAIC approach may not work well for some types of outcomes requiring a nonlinear model.  This issue can be mitigated by model-assisted (MA) CE (Wu \& Sitte, 2010), which balances the estimated outcome-covariate functions.  For example, for binary outcomes, one can fit a logistic regression for the outcome and covariates in one treatment group and balance the predicted probabilities between the two groups.  However, it is more difficult to apply this approach to time-to-event (TTE) data, due to the problem of censoring the time-dependent outcome-covariate relationship.

For using TTE data for indirect comparison, an extra challenge is to specify the right causal estimand. The hazard ratio (HR) in a Cox model (Cox, 1972) is often a problematic one, since it may not have a causal interpretation even without confounding biases (Hernán, 2010).   The use of HR for TTE with a cure fraction is more difficult to justify, since the proportional hazard assumption does not hold in the cured fraction part of the model at least.  For the analysis of TTE data with a cure fraction, several model-based approaches have been proposed.  A commonly used one is the mixture cure model (Amico \& Keilegom, 2018) with two components: the proportion for the cured  subjects, and a proper survival function for the events from the remaining subjects. Obviously, the model based approach requires correctly specified models for both components, since the two model components cannot be fitted separately.  

We propose a simple, intuitive and efficient approach to indirect comparison and borrowing with time-to-event data using CE approaches. To avoid dealing with censoring directly, we take the pseudo observation (PO) approach (Anderson et al., 2007, 2017; Klein et al., 2007) so that some simple weighting approaches such as MAIC can be applied to it directly. Here, we concentrate on MAIC, but our approach can also use other CE methods. 

The paper is organized as follows. The next section gives a brief review of methods we used for our approach, including the Neyman-Rubin framework (Rubin, 1974), CE approaches, in particular, MAIC, and POs for TTE data analysis. Models for data with a cure fraction are introduced in Section 3, where we show that one can obtain the PO of cure fraction from the PO of the survival function, and  propose our approach using POs and covariate balancing to the estimation of treatment effect in cure rate.   In Section 4, a simulation is conducted to examine different estimators and to compare with the commonly used IPW estimator. We also present an application of the approach, in Section 5, for the estimation of the cure rate of control treatment in the breast cancer cohort GSE6532. Finally, we end with some discussion points in Section 6. Technical details including R-codes for the application can be found in the Appendix. 

\section{Cure rate and survival function estimation label{ate}}
We denote the survival,cumulative hazard and distribution functions of the event time $T$ under treatment $D=d, d=0, 1$ for control and treated, respectively, as $S_d(t), H_d(t)$ and $F_d(t)$, respectively.  Their conditional counterparts given the covariates $\bX$ are denoted by  $S_d(t|\bX), H_d(t|\bX)$ and $F_d(t|\bX)$, respectively.
We are interested in treatment effect in the survival function and cure rate, in particular when the latter is non-zero. In this situation, $F_d(t)$ is a sub-distribution (Fine \& Gray, 1999).  But we will drop "sub" hereafter for simplicity, unless explicit distinction is needed.    As stated above, those for the treated subjects are easy to estimate from the trial data, and the challenge is to estimate $S_0(t)$ and $q_0=S_0(\infty)$: the survival function and the cure rate under control treatment using historical control data. 

We define formally the cure rate under the control treatment as $S_0(\infty)$.  A non-zero cure rate is often indicated when the survival curve has a flat tail at the end of observation.    
One commonly used model  to accommodate a fraction of cured subjects is the mixture cure model (Amico \& Keilegom, 2018):  
    \begin{equation}
        S_0(t|\bX_i)=1-p_0(\bX_i) +p_0(\bX_i) S_0^u(t|\bX_i),
        \label{mixcure0}
    \end{equation}
    where $1-p_0(\bX_i)$ is the cure fraction and $S_0^u(t|\bX_i)$ is a proper survival function for subjects who are not cured. 
    Models for $p_0(\bX_i)$ and $ S_0^u(t|\bX_i)$ will be referred to as the P-part and S-part of \eqref{mixcure0}, respectively. One can use model (1) for population adjustment, however, the two parts of this model have to be fitted simultaneously, hence the adjustment is sensitive to misspecification of the two parts. 
    Our goal is to estimate the control cure rate $q_0=E(1-p_0(\bX_i))$ in the trial population using external data. This can be done by fitting model \eqref{mixcure0} then using the P-part, but it is sensitive to model misspecification.  Therefore, we aim at developing alternatives independent of the cure model. 

Another commonly used model to accommodate a fraction of cured subjects is the promotion time cure model (Amico \& Keilegom, 2018):  
    \begin{equation}
        S_0(t|\bX_i)=\exp( -\theta_0(\bX_i) F_0^u(t|\bX_i)),
        \label{ptc}
    \end{equation}
where $\theta(\bX_i)$ is a non-negative function of $\bX_i$ and $F_0^u(t|\bX_i)$ is the distribution function corresponding to $S_0^u(t|\bX_i)$.  With this model, the individual cure fraction equivalent to $1-p_0(\bX_i)$ in the mixture cure model is $\exp( -\theta_0(\bX_i))$. One can also model the $\theta(\bX_i)$ and $F_0^u(t|\bX_i)$ parts, but they also have to be fitted simultaneously.  

\section{An pseudo observations approach with covariate balancing adjustment}
\subsection{Pseudo observations for time to events}
In this section, we propose our approaches to the estimation of the cure rate and survival function using POs and MAIC weights.  These approaches are model independent, and the cure rate can be estimated independently. As stated in Section 2, we will concentrate on the estimation of cure rate and survival function of control, adjusted to the trial population. Both approaches depend on the use of POs, which are briefly introduced as follows.

Censoring in survival analysis presents additional challenges in borrowing historical controls.    
The use of POs is a simple approach to survival analysis to avoid dealing with censoring (Andersen et al., 2017).  In our context, the PO for the survival function of subject $i$ of historical control  is defined as
\beqn
\hat S_0^i(t)=n \hat{S}_0(t) - (n-1) \hat{S}_0^{-i}(t),
\eeqn
where $\hat S_0(t)$ and $\hat S_0^{-i}(t)$ are the Kaplan-Meier estimators with and without subject $i$, respectively, based on historical controls.  
One can show that (Graw et al., 2009) 
\beqn
E(\hat S_0^i(t)|\bX_i) \approx S_0(t|\bX_i)
\label{graw}
\eeqn
so that $\hat S_0^i(t)$ can be treated as an observed individual survival rate at time $t$ in further analysis. Therefore, one can plug in $\hat S_0^i(t)$ as $Y_i$ to estimate $S_0(t)$ at a given time $t$ using \eqref{mu}. In the next section, we show that POs for the cure rate can also be obtained easily. 

\subsection{Cure rate estimation with covariate balancing adjustment \label{seccurerate}}
Our approach is to separate the P- and S-parts so that we can estimate the former independently. 
    The following property provides an easy way to population adjustment for cure rate using pseudo observations.  

{\bf Proposition 1:}
Suppose that 1) the extreme times satisfy $\tau_F  \le \tau_C$ where $\tau_F=\sup_t(t: F(t|x)<1)$ and $\tau_C$ is defined in the same way by the censoring distribution; 2) Censoring is independent of $T_i$ and $\bX_i$; 3) $F_0(t|\bX_i)$ is continuous at $\tau_C$ when $\tau_C< \infty$. Then, when $n \rightarrow \infty$  
\begin{equation}
  E(\hat S_0^i(T_n)|\bX_i)=1-p_0(\bX_i)+o(1),  
  \label{curepo}
  \end{equation}
where $T_n$ is the last event or censored time in the historical data. 

An outline of the proof is given in the Appendix.
Intuitively, this property holds because of the property of sub-survival function at the tail (Maller, 1992) and the property of $\hat S_0^i(t)$ given in \eqref{graw}. In practical aspect, the three conditions depend on sufficient follow-up and are indicated by a flat tail of the survival function.    

With this property, we can estimate $q_0$ using the pseudo observations of the cure rate $\hat S_0^i(T_n)$. The CE  approach (Deville \& S\"arndal, 1992) can be used to adjust for the population difference between the trial and the historical data.  The CE approach, adapted for our context, is to find weights $w_i$ for the historical control subjects such that  
\begin{align}
 \sum_{i=1}^n w_i \bX_i=\bar{\bX}^t ,
    \label{ce}
\end{align}
where $\bar{\bX}^t$ is the sample mean of $\bX_j$ in the trial population. Then the outcome data, the POs of the cure rate and survival function in our context, are weighted with the same weights before comparison with the trial data.   
Specifically, we propose to estimate the cure rate in the trial population by
\begin{equation}
    \hat q_0^{po}=\sum_{i=1}^{n} w_i \hat S_0^i(T_n),
    \label{q0po}
\end{equation}
The details of CE, including how to use models for better covariate balance, are given in the appendix.  

One problem with using estimator \eqref{q0po} is that $\hat S_0^i(T_n)$ may not be in the range of [0, 1].  This is a common problem of pseudo observations based estimator, including the unadjusted mean cure rate. An approach to solve this problem is to follow a similar approach as in Andersen and Klein (2007) and fit an estimating equation for a pseudo logistic model
\begin{equation}
    \sum_{i=1}^n w_i [\hat S_0^i(T_n) - 1/(1+\exp(-b))]=0.
\end{equation}
and obtain estimate $\hat b$. Then we can estimate
\begin{equation}
    \hat q_0^{pol}=1/(1+\exp(-\hat b)).
    \label{q0pol}
\end{equation}
One can also obtain a confidence interval for $b$, then convert it to the CI for $q_0$ with the same function so that the CI is also within [0,1].

For statistical inference based on $\hat q_0^{po}$, one can estimate its variance with existing formulae, then utilize an approximate normal distribution, or use the bootstrap approach.  The details of variance estimation are given in the Appendix.    Our experience prefers the latter, either with a small bootstrap runs to estimate the variance, or with a large number of runs to obtain the confidence interval directly.   
The above weighting approach can be used to estimate the survival function $S_0(t)$ in the trial population at a given time $t$. For this purpose, one can use  pseudo observations $\hat S^i(t)$ in the estimator
\begin{equation}
    \hat S_0^{po}(t)=\sum_{i=1}^{n} w_i \hat S_0^i(t).
    \label{posurv}
\end{equation}
Although $\hat S_0^{po}(t)$ at the end may be similar to $\hat q_0^{po}$, the latter is estimated based $\hat S_0^i(T_n)$ with random $T_n$.    

\subsection{Estimation based on weighted Kaplan-Meier estimator}
Another approach of using the MAIC weights is to weight the Kaplan-Meier estimator rather than the POs.  It may not be clear on how balancing eliminates confounding bias due to population difference. In fact, its validation depends on the doubly robust property described in Section \ref{appce}, because $w_i$ can also be considered as the inverse probability weight.  Hence, we can use $w_i$ in  the approach of Xie (2005). Let $t_j$ be the $j$th event time and $s_{0j}=S_0(t_j)/S_0(t_{j-1})$. The latter can be estimated by
\begin{equation}
    \hat s_{0j}=(1-\frac{\sum_{i: T_i=t_j} w_i \delta_i}{\sum_{i: T_i=t_j} w_i}),
    \nn
\end{equation}
where $\delta_i=1$ if the $i$th subject has an event at time $t_j$, and $\delta_i=0$ if censored. Following the proof of Eq 3 in Xie (2005), we conclude that $\hat s_{0j}$ is indeed a consistent estimator for $s_{0j}$. Then, $S_0(t)$ can be estimated by
\begin{equation}
    \hat S_0^{km}(t) =\prod_{j'=1}^j \hat s_{0j'}, \:\: t_j \le t \le t_{j+1}.
    \label{kmwei}
\end{equation}
 This method is different from \eqref{posurv}, as weighting is applied in the Kaplan-Meier estimator directly. 
 
 This leads to another approach to estimating $q_0$: given  weighted KM estimator $\hat S_0^{km}(t)$ in \eqref{kmwei}, $q_0$ can be estimated by 
$\hat S_0^{km}(T_n)$. This approach can be justified by the property of sub-survival function at the tail (Maller, 1992).  However, $\hat q_0^{po}$ allows the use of a model specifically for the cure rate, while the adjustment for the whole $S_0(t)$ may not be specific for its tail at $T_n$.

\section{A simulation study \label{simu}}
The purpose of this simulation study is mainly to evaluate the performance of the proposed estimators for cure fraction $q_0$, but the MAIC and MA approaches for estimating the survival function $S_0(t)$ using POs are also examined. The simulation for $q_0$ estimation compares PO based estimators with MAIC, MA and MAIC+MA, respectively.  For the last one, the KM based estimator $\hat S_0^{km}(T_n)$ is also compared. The part for $S_0(t)$ estimation compares PO based estimators with MAIC, MA. The comparison also includes the IPW KM approach, since it is commonly used for survival function.    

The simulation settings for the estimation of both $q_0$ and $S_0(t)$ are similar and are described as follows: 
\begin{enumerate}
    \item $\bX_i \sim N({\bf m},\bI_p)$ with means ${\bf m}$ and $\bI_p$ is a $p$-dimensional identity matrix.  The covariates means  $\bar \bX^t$, $q_0$ and $S_0(t)$ of the target (trial) population are generated from a random sample of 100,000 subjects.   The purpose is to obtain approximate true values of $q_0$ and $S_0(t)$ for comparison, and also set $\bar \bX^t$ as a target to match. In practice, the latter should be replaced by the sample mean in the trial population. 
    \item The time-to-event data are generated from the Weibull distribution with the scale parameter $\exp(3+\bX_i^T \boldsymbol{\gamma})$ with $\boldsymbol{\gamma}= g \bf 1$, and the shape parameters 0.5,  1 (exponential distribution) and 2. The censoring time is generated from an exponential distribution with scale $\exp(5.5)$.
    \item The cure status is generated from a binary distribution with cure rate $1/(1+\exp(-\bX_i \boldsymbol{\alpha}))$ with $\boldsymbol{\alpha}=a \bf 1$.
    \item The base scenario of the simulation used $n=200$, $p=3$, ${\bf m}=b \bone$,  and $b=0.5$, which gives censoring rates around 30\%. Other scenarios are  created with varying values of these parameters. 
\end{enumerate}
For each scenario, 2000 simulation runs are generated.  The Appendix gives details about the implementation of these weight calculation algorithms. Estimators evaluated in the simulation are MAIC, MA and MAIC+MA, which balance covariates, model prediction and both, respectively. For the MA and MAIC estimators, a logistic model is fitted to obtain model predictions. To evaluate the difference between using PO and KM, an additional estimator included in the simulation is the KM estimator based on the MAIC+MA weights. The biases of unadjusted estimates are also presented for reference.  

Based on the simulation result, the (100 $\times$) bias and standard error (SE) of the above estimators are presented in Table \ref{biasse}. All estimators significantly reduce the bias, although small biases up to around 3\% occur in different settings.  These are probably attributed to small sample sizes, since the bias reduces with the increase in the sample size or with a smaller $p$, the number of covariates, in the same setting. As expected, the SE generally decreases with either sample size increase or with a small $p$. The MA estimator generally has a lower SE than that of MAIC when the sample size is large. The MAIC+MA estimator has higher SE in most settings, suggesting that although the estimator has some desired theoretical properties, its small sample performance may be inferior to others. Finally, the weighted KM-based estimator $\hat S_0^{km}(T_n)$ behaves similar to the PO based one, in terms of both bias and SE.     

\begin{figure}
    \centering
    \includegraphics[width=140mm]{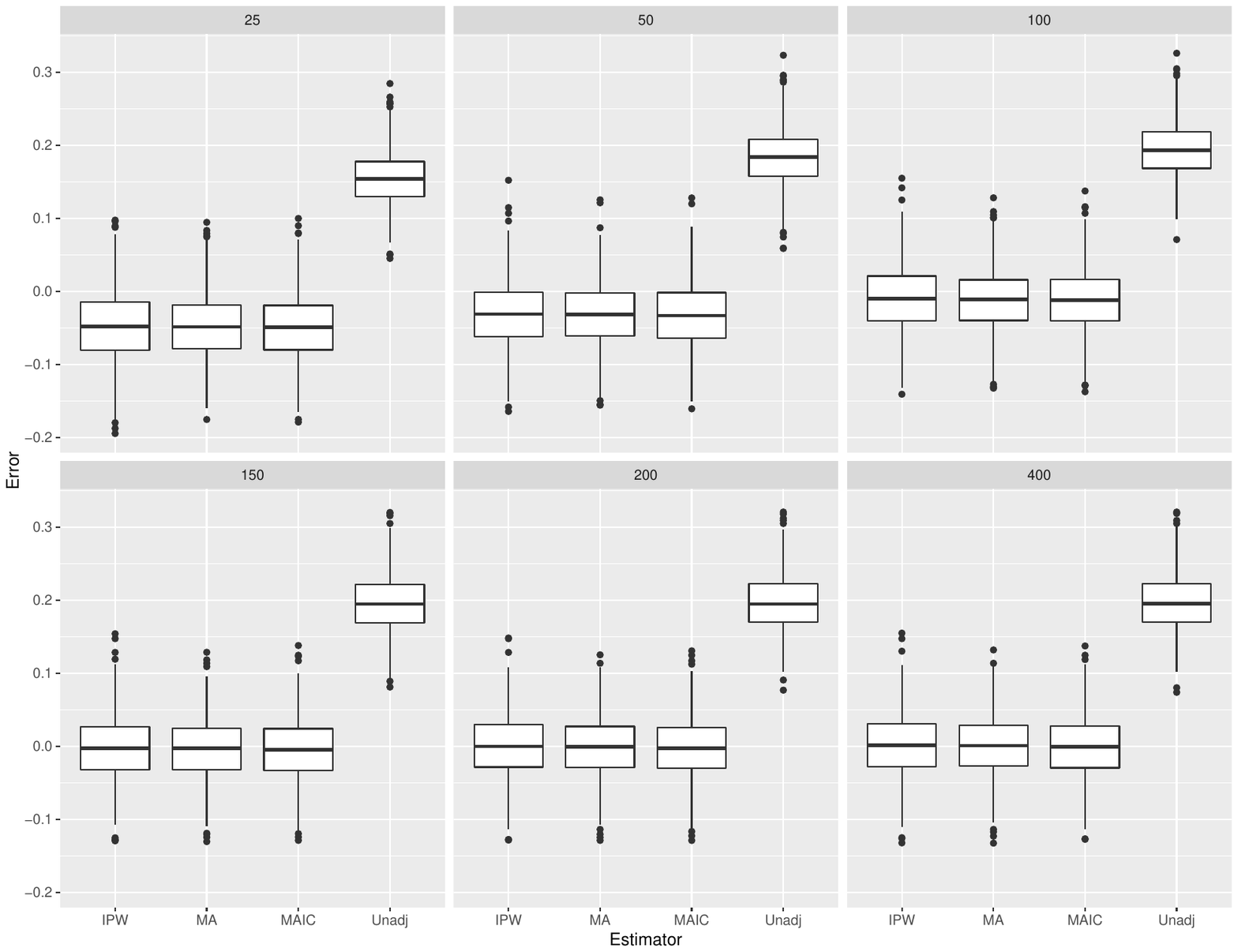}
    \caption{Error distribution of the IPW, MAIC and MA weighted estimators for $S_0(t)$ based on POs, compared with the unadjusted KM estimator at times 25, 50, 100, 150, 250 and 400.}
    \label{fig0}
\end{figure}

As an additional investigation, we also conduct a small simulation to examine the same PO based approach  to the estimation of survival function in the presence of a cure rate. The simulation setting takes that for the first block in Table \ref{biasse}, with sample size $n_1=300$. Survival function is estimated at $t=25, 50, 100, 150, 250$ and 400. Based on the simulation results in Table \ref{biasse}, we concentrate on the MAIC and MA estimators, compared with the IPW KM estimator.    
Figure \ref{fig0} presents the error distribution of the IPW, MAIC, and MA calibration estimators for the survival function at specific time points, together with the unadjusted estimator for reference, and shows an upward bias 15\% to 20\% at all time points.
A small downward bias occurs in early time points for all methods, but the bias reduces to zero at later time points.  All three estimators perform similar, although the IPW estimator generally has more extreme values.  The MA approach has slightly lower variability than the others, which is confirmed by a smaller variance of around 7\%, compared with that of MAIC.  

\begin{table}[ht]
    \caption{Bias and SE of estimated cure rate $q_0$ using MAIC, MA,MAIC+MA weighted PO, and MAIC+MA weighted KM. \label{biasse}}
\centering
\small
\begin{tabular}{rrrr|c|cccc|cccc}
  \hline
    & & &  & \multicolumn{5}{c}{Bias$\times$100}  & \multicolumn{4}{c}{SE $\times$100}\\

$n$  & $a$ & $g$  & $p$ & Unadj & MAIC & MA & \multicolumn{2}{c}{MAIC+MA } & MAIC & MA & \multicolumn{2}{c}{MAIC+MA } \\ 
    &&&&&PO& PO &PO& KM &PO& PO &PO& KM\\
\hline

  \hline
100 & 0.70 & -0.3 & 3 & -14.5 & 1.5 & 1.1 & 1.7 & 1.8 & 6.8 & 5.0 & 14.7 & 14.4 \\ 
200 & 0.70 & -0.3 & 3 & -15.0 & 0.7 & 0.8 & 1.1 & 1.3 & 4.7 & 2.7 & 9.3 & 9.4 \\ 
500 & 0.70 & -0.3 & 3 & -15.1 & 0.2 & 0.3 & 0.2 & 0.2 & 2.9 & 1.6 & 3.2 & 3.2 \\ 
200 & 0.70 & -0.3 & 5 & -20.7 & 1.9 & 0.1 & 1.2 & 1.2 & 5.9 & 5.6 & 10.8 & 10.3 \\ 
500 & 0.70 & -0.3 & 5 & -21.1 & 1.0 & 0.6 & 0.5 & 0.5 & 3.9 & 2.3 & 4.5 & 4.3 \\ 
\hline
100 & 0.70 & 0.3 & 3 & -13.6 & -0.4 & 1.8 & 2.6 & 2.0 & 7.3 & 8.0 & 16.3 & 14.9 \\ 
200 & 0.70 & 0.3 & 3 & -14.4 & -0.5 & 1.5 & 1.5 & 1.3 & 5.0 & 4.0 & 10.7 & 9.8 \\ 
500 & 0.70 & 0.3 & 3 & -14.7 & -0.5 & 0.6 & 0.5 & 0.5 & 3.3 & 2.1 & 4.2 & 4.2 \\ 
200 & 0.70 & 0.3 & 5 & -19.3 & -0.2 & 1.6 & 3.3 & 2.3 & 7.5 & 9.9 & 14.1 & 12.1 \\ 
 500 & 0.70 & 0.3 & 5 & -20.1 & -0.2 & 1.7 & 1.1 & 0.7 & 4.9 & 4.2 & 7.0 & 5.7 \\ 
 \hline
 100 & 0.35 & 0.3 & 3 & -7.7 & 0.1 & 1.7 & 2.8 & 3.2 & 7.9 & 9.0 & 22.7 & 21.4 \\ 
 200 & 0.35 & 0.3 & 3 & -8.7 & -0.6 & 1.3 & 1.9 & 1.7 & 5.1 & 7.7 & 16.7 & 15.3 \\ 
 500 & 0.35 & 0.3 & 3 & -9.1 & -1.1 & 0.4 & 0.6 & 0.5 & 2.7 & 2.7 & 9.6 & 9.0 \\ 
 200 & 0.35 & 0.3 & 5 & -11.5 & 0.4 & 2.8 & 5.2 & 4.2 & 7.8 & 9.3 & 26.1 & 22.2 \\ 
 500 & 0.35 & 0.3 & 5 & -12.4 & 0.2 & 1.7 & 3.0 & 2.5 & 4.5 & 4.9 & 19.0 & 16.3 \\ 

 \hline
 \multicolumn{13}{c}{Shape=2} \\
   \hline
100 & 0.70 & -0.3 & 3 & -14.7 & 2.3 & 1.0 & 2.9 & 2.8 & 6.6 & 4.6 & 15.7 & 15.0 \\ 
200 & 0.70 & -0.3 & 3 & -15.1 & 0.9 & 1.0 & 1.4 & 1.5 & 4.5 & 2.5 & 8.9 & 8.9 \\ 
500 & 0.70 & -0.3 & 3 & -15.1 & 0.4 & 0.6 & 0.3 & 0.3 & 2.9 & 1.5 & 3.3 & 3.0 \\ 
200 & 0.70 & -0.3 & 5 & -20.8 & 2.6 & 0.4 & 1.6 & 1.6 & 5.6 & 5.6 & 10.0 & 9.7 \\ 
500 & 0.70 & -0.3 & 5 & -21.1 & 1.3 & 1.0 & 0.6 & 0.6 & 3.7 & 2.1 & 2.8 & 2.7 \\ 
\hline
100 & 0.70 & 0.3 & 3 & -14.1 & -0.7 & 1.9 & 3.1 & 2.8 & 6.8 & 7.1 & 15.3 & 14.2 \\ 
200 & 0.70 & 0.3 & 3 & -14.6 & -0.6 & 1.3 & 1.5 & 1.3 & 4.8 & 3.9 & 9.8 & 9.1 \\ 
500 & 0.70 & 0.3 & 3 & -14.9 & -0.6 & 0.4 & 0.4 & 0.4 & 3.2 & 2.0 & 4.2 & 4.1 \\ 
200 & 0.70 & 0.3 & 5 & -20.0 & -0.3 & 1.2 & 2.5 & 1.7 & 7.2 & 9.4 & 11.9 & 9.7 \\ 
 500 & 0.70 & 0.3 & 5 & -20.6 & -0.2 & 1.3 & 0.6 & 0.4 & 4.7 & 3.7 & 4.5 & 4.3 \\ 
 \hline
 100 & 0.35 & 0.3 & 3 & -8.3 & -0.2 & 2.3 & 3.3 & 3.5 & 7.6 & 9.3 & 21.3 & 20.3 \\ 
 200 & 0.35 & 0.3 & 3 & -9.1 & -1.0 & 1.1 & 2.0 & 1.9 & 4.9 & 9.8 & 16.9 & 15.9 \\ 
 500 & 0.35 & 0.3 & 3 & -9.3 & -1.2 & 0.4 & 1.3 & 1.2 & 2.5 & 2.5 & 8.5 & 8.5 \\ 
 200 & 0.35 & 0.3 & 5 & -12.2 & 0.1 & 2.2 & 3.5 & 3.2 & 7.5 & 8.2 & 23.3 & 21.2 \\ 
 500 & 0.35 & 0.3 & 5 & -12.9 & 0.0 & 1.2 & 1.2 & 1.2 & 4.4 & 4.6 & 16.4 & 14.9 \\ 
\hline
 \multicolumn{13}{c}{Shape=1/2} \\
   \hline
100 & 0.70 & -0.3 & 3 & -13.1 & 2.6 & 1.1 & 4.2 & 4.2 & 7.6 & 5.8 & 21.2 & 19.9 \\ 
200 & 0.70 & -0.3 & 3 & -14.2 & 1.2 & 0.8 & 2.6 & 2.4 & 5.5 & 3.3 & 16.9 & 15.5 \\ 
500 & 0.70 & -0.3 & 3 & -14.7 & 0.5 & 0.2 & 0.3 & 0.3 & 3.4 & 1.9 & 7.3 & 5.9 \\ 
200 & 0.70 & -0.3 & 5 & -20.0 & 2.3 & 0.2 & 1.8 & 1.8 & 6.6 & 6.2 & 13.6 & 13.3 \\ 
500 & 0.70 & -0.3 & 5 & -20.6 & 1.3 & 0.6 & 0.6 & 0.4 & 4.4 & 2.8 & 7.8 & 7.2 \\ 
\hline
100 & 0.70 & 0.3 & 3 & -10.2 & 2.1 & 2.3 & 5.2 & 5.0 & 8.9 & 9.2 & 26.6 & 23.1 \\ 
200 & 0.70 & 0.3 & 3 & -12.1 & 0.6 & 2.4 & 2.7 & 2.5 & 6.1 & 7.3 & 18.1 & 16.0 \\ 
500 & 0.70 & 0.3 & 3 & -13.0 & 0.1 & 1.3 & 1.6 & 1.4 & 3.8 & 2.8 & 10.2 & 8.7 \\ 
200 & 0.70 & 0.3 & 5 & -15.5 & 1.3 & 3.3 & 5.3 & 4.0 & 8.6 & 9.9 & 22.6 & 19.0 \\ 
 500 & 0.70 & 0.3 & 5 & -17.2 & 0.8 & 3.5 & 3.1 & 2.7 & 5.8 & 6.3 & 15.2 & 11.6 \\ 
 \hline
 100 & 0.35 & 0.3 & 3 & -4.1 & 3.1 & 1.6 & 5.3 & 6.4 & 9.8 & 10.5 & 31.8 & 28.6 \\ 
 200 & 0.35 & 0.3 & 3 & -6.3 & 1.1 & 1.0 & 4.5 & 4.7 & 6.9 & 8.6 & 24.2 & 22.9 \\ 
 500 & 0.35 & 0.3 & 3 & -7.6 & -0.2 & 0.7 & 1.7 & 1.5 & 3.5 & 5.4 & 15.2 & 14.0 \\ 
 200 & 0.35 & 0.3 & 5 & -7.8 & 1.8 & 1.5 & 6.0 & 5.9 & 8.6 & 6.3 & 31.9 & 26.7 \\ 
 500 & 0.35 & 0.3 & 5 & -9.6 & 1.2 & 1.7 & 4.2 & 3.6 & 5.3 & 6.2 & 25.2 & 20.7 \\ 

 \hline
\end{tabular}
\end{table}

\section{Estimation of cure rate for control group in breast cancer cohort GSE6532 \label{app}}
\begin{figure}
    \centering
    \includegraphics[width=120mm]{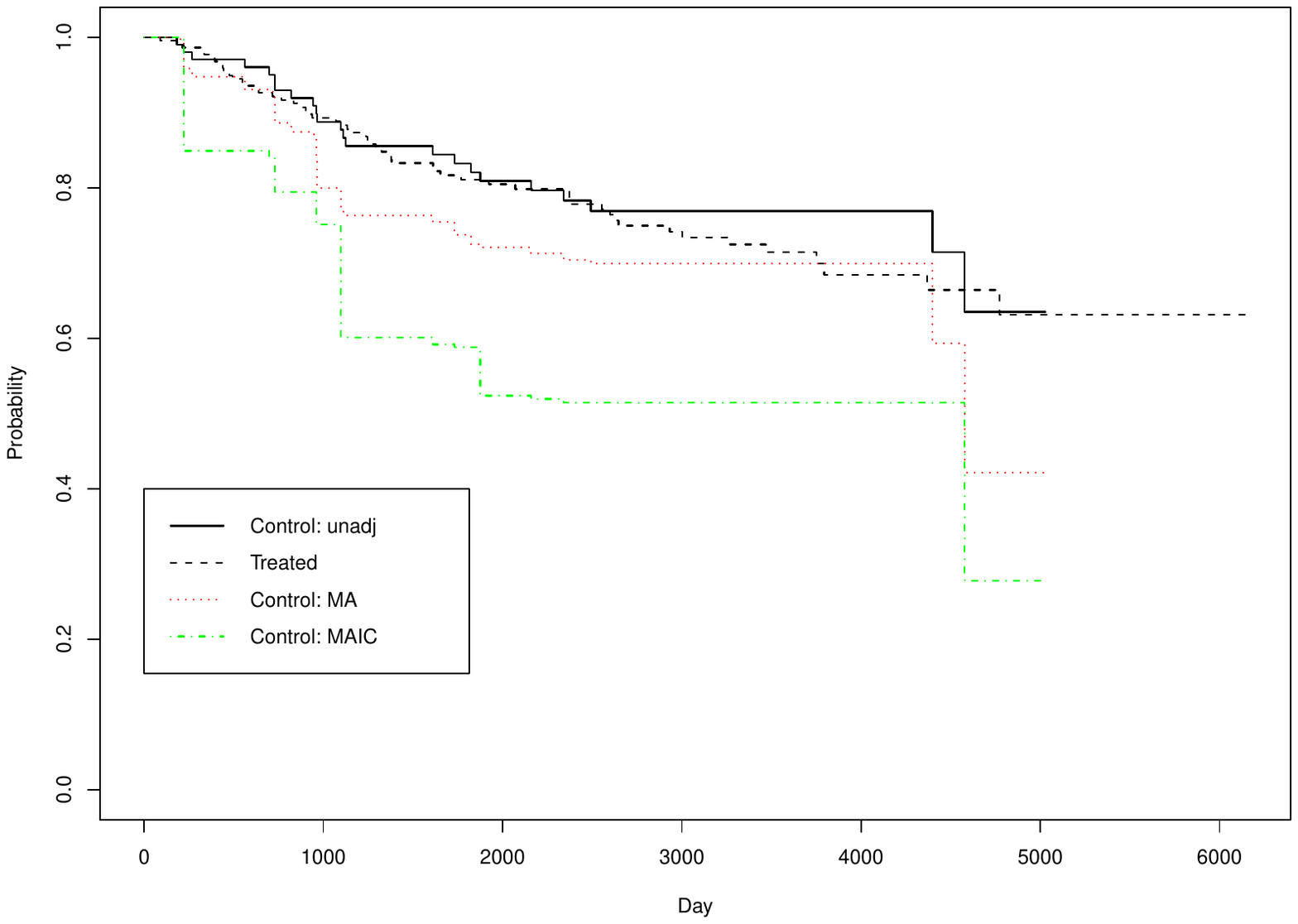}
    \caption{Kaplan-Meieir curves of distant metastasis-free surviva for subjects treated with Tamoxifen, compared with control curves of unadjusted, MAIC and MA weighted, respectively. }
    \label{figintro}
\end{figure}

The proposed approaches are applied to the cohort GSE6532 in data collected by Loi et al. (2007) to compare Tamoxifen with
 a control for treating patients with breast carcinomas.   Treatments were not randomized, hence we used MAIC to adjust for potential biases.  The dataset includes baseline covariates age, tumour grade and size, node involvement, and estrogen receptor (positive/negative). Some patients have missing values in some covariates, hence are excluded from the analysis.  The complete dataset has 219 and 104 patients in the Tamoxifen  and the control groups, respectively. Node involvement is not used, as no patient in the control group has node involvement. The mean or percentage difference between Tamoxifen treated and control patients are 12.8 years, 13\%, -6\%, 0.43 and 28\% for age, tumour grades 2 and 3, tumour size and estrogen receptor positive, respectively.  These differences indicate that Tamoxifen treated patients might have worse diagnosis.
 Distant metastasis-free survival curves and cure rates of Tamoxifen treated, and control patients seemed similar, as presented in Figure \ref{figintro}.  
Although the survival curves are not long enough to confirm non-zero cure rates,  we use this dataset  for illustration purposes, due to the lack of more appropriate real data that are publishable. 
 
 Suppose that we are interested in the estimation of cure rate of distant metastasis-free survival, which is one main outcome of interest in the study.   However, control patients might have a better prognosis.  As previously shown, the challenge is the estimation of the cure rate and survival curve under control treatment, but for Tamoxifen treated subjects; hence we will concentrate on this task. Based on the simulation result in the previous section, we
 concentrate on the MAIC and MA approaches and compare them with the direct ANCOVA adjustment estimator, all based on pseudo-observations, to estimate this cure rate. 
 
\begin{figure}
    \centering
    \includegraphics[width=130mm]{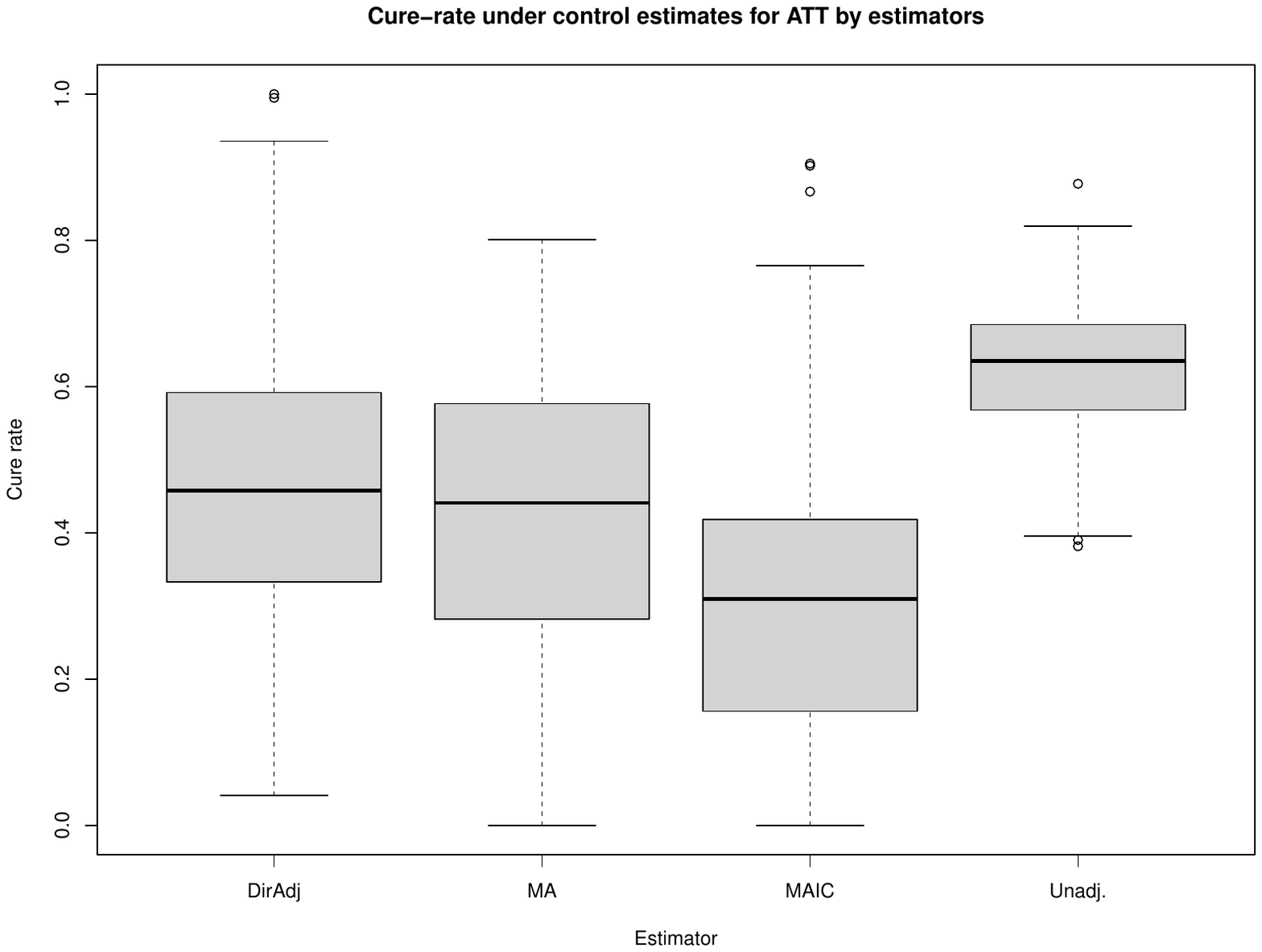}
    \caption{Distribution of cure rate estimates under control for Tamoxifen treated population in GSE6532, direct ANCOVA adjustment, MA and MAIC weighted PO estimators, compared with the unadjusted one.  The estimates are based on pseudo logistic models.}
    \label{tamlogit}
\end{figure}

 To overcome the problem of negative estimates or confidence intervals, we applied the pseudo logistic model approach to obtain estimates.  
 The estimated cure rates are 0.28 by MAIC balancing all the above covariates; 0.44 by model-based calibration balancing model prediction based on a fitted logistic model fitting pseudo observation $\hat S^{i}(T_n)$ and covariates data; and 0.46 by direct adjustment using the same fitted model. For comparison, the unadjusted estimate is 0.64.  
 The distribution of the estimates with 300 bootstrap runs is presented in Figure \ref{tamlogit}. As expected, model-assisted calibration and direct adjustment give similar results, although the former has slightly larger variability. The MAIC approach generally has higher variability than the other two estimators, with considerable mean difference.  
 
  The MA and MAIC weights were also applied to the Kaplan-Meier estimator for the controls to adjust it to the Tamoxifen treated population. The adjusted control curves are also presented in Figure \ref{figintro}, which are consistent with the adjusted cure rates, and also indicate earlier differences between Tamoxifen treated and control curves. 

We also applied the R-package smcure (Cai et al., 2022), which implements the EM algorithm for mixture models. The same covariates are included in both the survival and cure rate parts. after fitting the model, the cure rate is estimated by the mean predicted cure rate of the Tamoxifen treated subjects using the P-part of (\ref{mixcure0}),  and results in an estimated cure rate of 0.32, which, interestingly, is similar to the MAIC estimate 0.28.   Although the package offers variance estimation via bootstrap, it failed to converge during some bootstrap runs for this data.  This suggests that the results may depend on model assumptions.  Nevertheless, all estimators results in much lower cure rates than the unadjusted one. 

\section{Discussion}
Survival analysis for data with a cure rate is challenging, especially for the estimation of causal effect. We propose estimation approaches based on pseudo observations and calibration estimation, in particular the MAIC approaches with and without model assistance.  We show that the pseudo observations of the survival function at the last event or censoring time can be considered as pseudo observations for the cure rate, hence MAIC and other CE approaches can be applied to them directly to estimate the cure rate. 

One open question is how to estimate the proper survival function $S^u_0(t)$, without specifying a mixed model. Although we can estimate $q_0$ separately and obtain the PO of $S_0(t|\bX_i)$, it seems not possible to separate the two terms in $p_0(\bX_i) S_0^u(t|\bX_i)$ without model specification.  Intuitively, the treatment may not cure the disease at a single time point, hence the process of cure may happen at a much earlier time than $T_n$ and cannot be nonparametrically identified. 

We have concentrated on entropy balancing; i.e., the MAIC approach for calibration in this paper.  Several other "distance" measures between weights as alternatives to the entropy balancing can also be used (Deville \& Sarndal, 1992). See Devaud and Till\'e (2019) for a review of recent developments in sample survey. 

The simulation result showed the difficulty of balancing multiple covariates, which led to worse performance of the corresponding estimators due to the extreme weights required by the constraints. This problem is similar to that of small $n$ and large $p$, but may occur even when $p$ is still numerically much smaller than $n$ (e.g., when $p=5$). It should be noted that the independence between elements in $\bX_i$ for the simulation is the worst scenario. A correlation between them makes balancing easier and improves the performance of the estimators (Wang, 2021).
An approach to mitigate the issue of extreme weights is to allow for a small imbalance. To this end, one can replace $\sum_{i=1}^n w_i \bX_i=\bar{\bX}^t$ with $|\sum_{i=1}^n w_i \bX_i-\bar{\bX}^t| \le \bf \delta$ with $\bf \delta$ determined by the importance of each covariate (Zubizarreta, 2015). 

The MA calibration estimator is consistent, but one may be uncomfortable with it as the outcome $Y_i$ is used twice, once to fit the outcome model, and again for the weighted mean \eqref{mu}.  This can be avoided by sample splitting MA calibration that uses a part of the data to fit the model and to calculate the weights, while the rest to calculate the weighted mean.   

Without being explicitly stated, our approach depends on the key assumption of no unobserved confounding factors,  like other approaches for indirect comparison and confounding bias adjustment, to fully eliminate the confounding bias. That is, the covariates  $\bX_i$ include all confounders.  This assumption has to be justified by clinical judgement and cannot be tested based on existing data. Nevertheless, the adjustment can at least reduce, if not fully eliminate, the bias.  When this approach is used to augment a small control group, it can be combined with Bayesian borrowing as an alternative to the approach utilizing propensity scores (Wang et al., 2019, 2020; Chen et al., 2020).  


\bibliographystyle{tfcad}


\newpage
\appendix
\section{Methodology details}
\subsection{Further details on calibration estimation \label{appce}}
\subsection{Calibration estimation for population adjustment \label{secce}}
 The idea of CE is to weight historical control subjects so that their weighted means of covariates match those in a target population.  Denote the outcome, covariates, and treatment of subject $i$ in the historical control population as $Y_i$, $\bX_i, i=1,...,n$, respectively. That is, $D_i=0, i=1,...,n$, but $\bX_i$ has different distribution from the covariates $\bX_j$ in the trial population.  To adjust for the difference,
 a general CE  approach (Deville \& S\"arndal, 1992), adapted for our context, is to find weights $w_i$ for the historical control subjects such that  
\begin{align}
 \sum_{i=1}^n w_i \bX_i=\bar{\bX}^t ,
    \label{gence}
\end{align}
where $\bar{\bX}^t$ is the sample mean of $\bX_j$ in the trial population. For efficiency, $w_i$ should be as close to uniform weight as possible, measured by a distance between $w_i$ and uniform weights $1/n$.  When the distance is the entropy $w_i \log(w_i)$ (Hainmueller, 2012), we have
\begin{equation}
    w_i=\exp(\bf \beta^T \bX_i)/\sum_{i=1}^{n} \exp(\bf \beta^T \bX_i).
    \label{maicwei}.
\end{equation}
Details on how $\bf \beta$ is calculated are given in \ref{appce}. 
With the weights defined above, we can estimate $\mu_0$ by
\begin{equation}
    \hat \mu_0= \sum_{i=1}^n w_i Y_i
    \label{mu}
\end{equation}
which is equivalent to the MAIC estimator (Signorovitch et al., 2012), and is referred to as such henceforth.To see how this approach work, it is easy to show that when $Y_i$ is a linear function of $\bX_i$, the weighting eliminates the impact of $\bX_i$ in $\hat \mu_0$.
Furthermore, the entropy balancing estimator is doubly robust (DR) in the sense that the estimator \eqref{mu} is consistent when either the outcome-covariate relationship or the logit of the propensity score is linear with respect to the covariates (Zhao \& Percival, 2017).   

This CE approach may not work well when the outcome model is nonlinear in nature, e.g., a logistic model for a binary outcome. One way of dealing with nonlinear outcome models is model-assisted (MA) CE (Wu et al. 2001). Let the outcome model be
    \begin{equation}
        E(Y_i(0)|\bX_i)=m(\bX_i^T \bf \gamma). 
        \label{ymodel}
    \end{equation} 
This approach first fits the model to historical control subjects to obtain $\hat \bf \gamma$. Then, the weights are determined to balance the predicted outcomes between the two populations, i.e.,
\begin{equation}
    \bar{\mbox{\bf m}}^t = \sum_{i=1}^n w_i m(\bX_i^T \hat \bf \beta),
    \label{mabalance}
\end{equation}
where $\bar{\mbox{\bf m}}^t$ is the mean of $m(\bX_j^T \hat \bf \gamma)$ in the trial population.  
Obviously, the validity of $\hat \mu_0$ depends on the model (\ref{ymodel}), which may be wrong. However, we can make $\hat \mu_0$ multi-robust by balancing multiple models in the same way as balancing multiple covariates in $\bX_i$. In addition, one can also combine $m(\bX_i^T \hat \bf \gamma)$  with $\bX_i$ to balance both by adding the conditions in \eqref{mabalance} into \eqref{gence}.  This MA approach together with MAIC balancing will be denoted as MAIC+MA. 

To see how the calibration works, suppose that the conditional mean under the treatment $D_i=0$ is $\mu_0(\bX_i)=\bf \gamma_0^T \bX_i$. Then, $E(\hat \mu_0) \approx \bf \gamma_0^T \sum_i^n w_i \bX_i=\bf \gamma_0^T \bar \bX^t$, which is exactly an estimate based on a fitted linear outcome model. However, the CE approach does not fit the model explicitly and does not need the outcome data, an important advantage in some situations. Although the justification only applies to linear outcome models, the estimator $\hat \mu_0$ is doubly robust in the sense that it is also valid if the logit of the propensity score is a linear function of $\bX_i$ (Zhao \& Pereival, 2017).  In this case, $w_i$ can be considered as the IPW weight.

\subsection{Technical details for Section \ref{seccurerate}}

To identify $p_0(\bX_i)$ we need $\tau_F  \le \tau_C$ where $\tau_F=\sup_t(t: F_0(t)<1)$ and $\tau_C=\sup_t(t: F_c(t)<1)$, where $F_c(t)<1)$ is the censoring distribution. Intuitively, to identify the cure rate, the right tail of $F_u(t)$ should not be censored.  
Even when the two parts are identifiable, maximizing  the marginal likelihood for the model (\ref{mixcure0}) is difficult, as it lacks a closed form and requires simulation. A more commonly used approach by
Peng and Dear (2000), and Sy and Taylor (2000), is based on the Expectation-Maximization (EM) algorithm (Dempster et al., 1977).  
Without considering $\bX_i$, Maller et al. (1992) showed that $\hat S_0(T_n)$ is a consistent estimator for the cure fraction $q_0$. This result plays a key role in our work and  motivated our approach of estimating $q_0$ independent of the S-part.  

Next we give an outline of proof of Proposition 1 is given without the full details, which can be found in Maller and Zhou (1992) and Graw et al. (2009).

{\bf Proposition 1}
Suppose that 1) the extreme times satisfy $\tau_F  \le \tau_C$ where $\tau_F=\sup_t(t: F(t|x)<1)$ and $\tau_C$ is defined in the same way by the censoring distribution; 2) Censoring is independent of $T_i$ and $\bX_i$; 3) $F_0(t|\bX_i)$ is continuous at $\tau_C$ when $\tau_C< \infty$. Then when $n \rightarrow \infty$  
\begin{equation}
  E(\hat S_0^i(T_n)|\bX_i)=1-p_0(\bX_i)+o(1)  
  \label{curepo1}
  \end{equation}
where $T_n$ is the last event or censored time in the historical population.

{\bf Outline of proof.}
This follows the same way as Maller and Zhou (1992, Theorem 1) which showed that the average cure rate can be estimated by the Kaplan-Meier estimate $\hat S_0(t)$ at $T_n$, using the fact that $T_n \le \tau_C$, but when $\tau_C \le \infty$, $T_n \rightarrow \tau_C$.  Using Lemma 2 in Graw et al. (2009), under assumptions 1) and 2) in Proposition 1, we have for any $t < \tau_C$,
$E(\hat S_0^i(t)|\bX_i) =S_0(t|\bX_i)+o(1)$.  Then  for $\tau_C < \infty$
\begin{align}
      E(\hat S_0^i(T_n)|\bX_i) &=S_0(T_n|\bX_i)+o(1)\nn\\
       &=S_0(\tau_C|\bX_i)+o(1)\nn\\
       &=1-p_0(\bX_i)+ p_0(\bX_i) S_0^u(\tau_C|\bX_i)+o(1)\nn\\
       &=1-p_0(\bX_i)+o(1),
\end{align}
where the second equality used the fact that $F_0(t|\bX_i)$ is continuous at $\tau_C$ when $\tau_C< \infty$, and the last one $\tau_F  \le \tau_C$ and consequently $S_0(\tau_C|\bX_i)=0$.  When $\tau_C = \infty$, $T_n \rightarrow \infty$ hence $S_0^u(\tau_C|\bX_i) \rightarrow 0$. 
Note that for any $\bX_i$ with $P(\bX_i>0)$, the assumption $\tau_F  \le \tau_C$ leads to $\tau_{F_x} =sup_t(t|F_0(t|\bX_i)<1) \le \tau_C$.  
    
\subsection{Statistical inference and variance estimation}
 Statistical inference for $\Delta$ based on any weighted estimator often depends on the estimation of its standard error (SE), or equivalently, its variance. As the variance of an estimate of $\mu_1$ can be estimated by its sample variance,  we will focus on the variance of $\hat \mu_0=\sum_{i=1}^n w_i Y_i$. 

Even with this simplification, estimating the variance of $\hat \mu_0=\sum_{i=1}^n w_i Y_i$ may still not be easy  in the general situation when only one of the outcome and propensity score models is correct, although $\hat \mu_0$ is doubly robust.  The original sandwich estimator proposed in (Signorovitch, 2012) seemed to be
\begin{equation}
  \hat V_0=\sum_{i=1} w_i^2 (Y_i- \hat \mu_0)^2.    
\end{equation}
The estimate $\hat V_0$ can be rather conservative, due to  the fact that the weights to balance the covariates are not taken into account.  Cheng et al (2020) derived a more accurate variance, assuming a correct propensity score model. The calculation is rather complex, even with some suggested simplifications.  
  A common variance estimator in survey sampling is 
\begin{equation}
 \hat V_{ss} =\sum_{i=1}^n w_i^2 (Y_i-\hat m(\bX_i))^2,
\end{equation}
where $\hat m(\bX_i)$ is a fitted outcome model.  This estimator does not have the issue of being conservative.  But it does depend on the fitted outcome model.  An approach not depending on fitted model has been proposed, but it may have a downward bias when the sample size is small or the number of covariates to be balanced is large (Wang, 2021).  

In practice, we propose using bootstrap to calculate either the confidence interval with a large number of bootstrap runs or the variance of the estimator with fewer bootstrap runs. Although numerical instability in simulation studies has been reported, it is rarely an issue in applications for a specific data set.  In the application (Section \ref{app}), we have illustrated the use of bootstrap to evaluate the distribution of cure fraction estimators.
\subsection{R-codes for Figures 2 and 3}
The R codes to produce Figures 2 and 3 illustrate how the proposed approaches are implemented in a real scenario.
\footnotesize
\begin{verbatim}
library(survival)
library(pseudo)
library(geepack)
library(smcure)

load("LUMINAL.RData")
all0=data.frame(rbind(demo.untreated,demo.tam),treat=rep(0:1,c(nrow(demo.untreated),nrow(demo.tam))))
inc=(all0$er !="NA" & all0$age !="NA" &all0$size !="NA"  & all0$e.dmfs !="NA" & all0$grade !="NA")
all1=all0[which(inc=="TRUE"),]

fn=function(bb){
  log(sum(exp(-Xi0%*%bb)))+sum(bb*mp1)
}
grr=function(bb){
  -t(Xi0)%*%exp(-Xi0%*%bb)/sum(exp(-Xi0%*%bb))+mp1
}
fnv=function(bb){
  log(sum(exp(-Xi0*bb)))+sum(bb*mp1)
}

all1=cbind(all1,logage=log(all1$age))

pd <- smcure(Surv(t.dmfs, e.dmfs)~treat+logage+as.factor(grade)+size+er,
             cureform=~treat+age+as.factor(grade)+size+er,
             data=all1,model="ph",Var = FALSE)
pd0 <- smcure(Surv(t.dmfs, e.dmfs)~age+grade+size+er,
             cureform=~age+as.factor(grade)+size+er,
             data=all1[all1$treat==0,],model="ph",Var = F)
treat=all1$treat
n1=sum(treat)
X1=cbind(rep(1,n1),all1[treat==1,c("age","grade","size","er")])
lp=as.matrix(X1)%*%pd0$b
mean(1/(1+exp(-lp)))

plot(survfit(Surv(t.dmfs, e.dmfs)~treat, data=all0),lty=1:2)

nall1=dim(all1)[1]

all=all1
treat=all$treat
#Mcov=model.matrix(treat~age+log(age)+as.factor(grade)+size+er,data=all)
Mcov=model.matrix(treat~age+as.factor(grade)+size+er,data=all)

#round(apply(Mcov[treat==1,],2,mean)-apply(Mcov[treat==0,],2,mean),2)

#  Figure 3

maxt=max(all$t.dmfs)
Yi=pseudosurv(time=all$t.dmfs, event=all$e.dmfs,tmax=maxt)$pseudo
Mcov00=Mcov[treat==0,]
Mcov1=Mcov[treat==1,]
Yi00=Yi[treat==0]
nsub=sum(treat==0)

Out=NULL
nsimu=300
for(simu in 1:nsimu){
  ind=sample(1:nsub,replace=TRUE)
  Yi0=Yi00[ind]
  Mcov0=Mcov00[ind,]
  mu1=mean(Yi[treat==1])
  Xi0=Mcov0[,-1]
  mp1=apply(Mcov1[,-1],2,mean)
  np=length(mp1)
  opt=try(optim(par=rep(0,np),fn=fn,gr=grr,method="BFGS", control=list(maxit=300, reltol=1e-8)))
  bb=opt$par
  wei=c(exp(-Xi0%*%bb)/sum(exp(-Xi0%*%bb)))
  fl=geese(Yi0~1,id=1:nsub, family="gaussian",mean.link="logit",weights = wei)$beta
  est=1/(1+exp(-fl))
  f0=geese(Yi0~1,id=1:nsub, family="gaussian",mean.link="logit")$beta
  unadj=1/(1+exp(-f0))
  beta=geese(Yi0~Xi0,id=1:nsub, family="gaussian",mean.link="logit")$beta
  pcure1=1/(1+exp(-Mcov1%*%beta))
  pcure0=1/(1+exp(-Mcov0%*%beta))
  Xi0=pcure0
  mp1=mean(pcure1)
  opt=optim(par=0,fn=fnv,method="BFGS", control=list(maxit=300, reltol=1e-8))
  bb=as.vector(opt$par)
  weima=exp(-pcure0*bb)/sum(exp(-pcure0*bb))
  fl=geese(Yi0~1,id=1:nsub, family="gaussian",mean.link="logit",weights = weima)$beta
  estma=1/(1+exp(-fl))
  estdir=mp1
  Xi0=cbind(pcure0,Mcov0[,-1])
  mp1=c(mean(pcure1),apply(Mcov1[,-1],2,mean))
  opt=optim(par=rep(0,np+1),fn=fn,gr=grr,method="BFGS", control=list(maxit=300, reltol=1e-8))
  bb=opt$par
  wei=c(exp(-Xi0%*%bb)/sum(exp(-Xi0%*%bb)))
  fl=geese(Yi0~1,id=1:nsub, family="gaussian",mean.link="logit",weights = wei)$beta
  estmama=1/(1+exp(-fl))
  Out=rbind(Out,c(unadj,est,estma,estmama,estdir))
}

round(apply(Out,2,mean),2)
apply(Out,2,var)

#Est=c(Out[,-1])
Est=c(Out[,-4])
Estimator=rep(c("Unadj.","MAIC","MA","DirAdj"),rep(nsimu,4))

boxplot(Est~Estimator, main="Cure-rate under control estimates for ATT by estimators", 
        xlab="Estimator", ylab="Cure rate")

#  Adjusted S_0(t) (Figure 2)
maxt=max(all$t.dmfs)
Yi=pseudosurv(time=all$t.dmfs, event=all$e.dmfs,tmax=maxt)$pseudo
Mcov00=Mcov[treat==0,]
Mcov1=Mcov[treat==1,]
Yi00=Yi[treat==0]
nsub=sum(treat==0)
  Yi0=Yi00
  Mcov0=Mcov00
  mu1=mean(Yi[treat==1])
  Xi0=Mcov0[,-1]
  mp1=apply(Mcov1[,-1],2,mean)
  np=length(mp1)
  opt=try(optim(par=rep(0,np),fn=fn,gr=grr,method="BFGS", control=list(maxit=300, reltol=1e-8)))
  bb=opt$par
  wei=c(exp(-Xi0%*%bb)/sum(exp(-Xi0%*%bb)))
  est=sum(Yi0*wei)
  unadj=mean(Yi0)
  beta=geese(Yi0~Xi0,id=1:nsub, family="gaussian",mean.link="logit")$beta
  pcure1=1/(1+exp(-Mcov1%*%beta))
  pcure0=1/(1+exp(-Mcov0%*%beta))
  Xi0=pcure0
  mp1=mean(pcure1)
  opt=optim(par=0,fn=fnv,method="BFGS", control=list(maxit=300, reltol=1e-8))
  bb=as.vector(opt$par)
  weima=exp(-pcure0*bb)/sum(exp(-pcure0*bb))
  estma=sum(Yi0*weima)
  par(lwd=1.5)
  plot(survfit(Surv(t.dmfs, e.dmfs)~treat, data=all1),lty=1:2,xlab="Day",ylab="Probability",lwd=1.5)
  Sma=survfit(Surv(t.dmfs, e.dmfs)~1, data=all1[treat==0,],weights=weima,conf.type = "none")
  Sce=survfit(Surv(t.dmfs, e.dmfs)~1, data=all1[treat==0,],weights=wei, conf.type = "none")
  lines(Sma,lty=3,col="red", conf.int = FALSE,lwd=1.5)
  lines(Sce,lty=4,col="green",conf.int = FALSE,lwd=1.5)
  legend(0,0.4,legend=c("Control: unadj","Treated","Control: MA","Control: MAIC"),lty=1:4,
         col=c("black","black","red","green"))
\end{verbatim}
\end{document}